\documentclass[twocolumn]{aastex63}

\newcommand{\Msun}{\ensuremath{M_{\odot}}}
\shorttitle{VLBA Observations of AT2019wey}
\shortauthors{Yadlapalli et al.}

\graphicspath{{./}{figures/}}
\usepackage{gensymb}
\usepackage{float}

\begin{document}

\title{VLBA discovery of a resolved source in the candidate black hole X-ray binary AT2019wey}

\email{nyadlapa@caltech.edu}

\author{Nitika Yadlapalli}
\affiliation{
Cahill Center for Astronomy and Astrophysics, California Institute of Technology, Pasadena, CA 91125, USA}

\author{Vikram Ravi}
\affiliation{
Cahill Center for Astronomy and Astrophysics, California Institute of Technology, Pasadena, CA 91125, USA}

\author{Yuhan Yao}
\affiliation{
Cahill Center for Astronomy and Astrophysics, California Institute of Technology, Pasadena, CA 91125, USA}

\author{S. R. Kulkarni}
\affiliation{
Cahill Center for Astronomy and Astrophysics, California Institute of Technology, Pasadena, CA 91125, USA}

\author{Walter Brisken}
\affiliation{
National Radio Astronomy Observatory, Soccoro, NM 87801, USA}

\begin{abstract}
AT2019wey is a Galactic low mass X-ray binary with a candidate black hole accretor first discovered as an optical transient by ATLAS in December 2019. It was then associated with an X-ray source discovered by SRG/eROSITA and SRG/ART-XC instruments in March 2020. After a brightening in X-rays in August 2020, VLA observations of the source revealed an optically thin spectrum that subsequently shifted to optically thick, as the source continued to brighten in the radio. This motivated us to observe AT2019wey with the VLBA. We found a resolved source that we interpret to be a steady compact jet, a feature associated with black hole X-ray binary systems in hard X-ray spectral states. The jet power is comparable to the accretion-disk X-ray luminosity. Here, we summarize the results from these observations.

\end{abstract}


\section{Introduction} \label{sec:intro}


Black hole X-ray binaries are comprised of stellar mass black holes accreting from companion stars. Most of the electromagetic emission from these systems arises from accretion disks and relativistic jets, and strong coupling is observed between properties of the disks and jets. The presence of a jet is dependent on which X-ray spectral state the black hole binary is observed to be in. This phenomenon was first observed in the correlated X-ray and radio intensities of Cyg X-1 \citep{Tananbaum1972}, where the X-ray emission was interpreted as originating from the disk, and the radio emission from the jet. The two main X-ray spectral states of interest are the thermal state (formerly known as the high/soft state) and the  hard state (formerly known as the low/hard state). In the thermal state, $>75$\% of the observed flux is contributed by the accretion disk. The predominantly thermal spectrum is accompanied by a steep power law extending to energies higher than \(\sim\)10\,keV. In the hard state the disk appears cooler and contributes very weakly, and $>80$\% of the flux is contributed by a non-thermal power law spectrum with a photon index of $1.4 < \Gamma < 2.1$. For detailed reviews, see \cite{Fender2003}, \cite{McClintock2009}, and \cite{Remillard2006}. 

The first observation of a resolved jet in an X-ray binary was conducted with the Very Large Array (VLA) on SS\,433 \citep{Hjellming1981}. Subsequent observations of the X-ray binary systems GRS\,1915+105 \citep{Mirabel1994} and GRO\,$1655-40$ \citep{Tingay1995}, with the VLA and the Southern Hemisphere VLBI Experiment (SHEVE) respectively, revealed the first Galactic examples of superluminal motion of relativistic jet components. The ejection of superluminal components was linked to X-ray outbursts. In its hard `plateau' X-ray state, however,  GRS\,1915+105 was observed to host a compact steady radio jet with mildly relativistic component velocities \citep{Dhawan2000}. Compact steady jets have also been observed in two other sources, Cyg\,X-1 \citep{Stirling2001} and MAXI\,J1836$-$194 \citep{Russell2015}, in canonical hard states. The link between the ejection of highly relativistic jet components and X-ray outbursts in the thermal state, first observed in GRS\,1915+105 \citep{Dhawan2000}, established a causal link between the accretion rate and jet properties \citep{Vadawale2003,fender04}. 

Radio observations of the hard-state compact steady jets of GRS\,1915+105 and Cyg\,X-1 that comfortably resolve the emission region were used to derive the jet powers and speeds \citep{Dhawan2000,Stirling2001}. These measurements are consistent with the model of conical synchrotron jets to describe radio emission from black hole accretors \citep{Blandford1979,Hjellming1988}, and motivated the derivation of scaling relations between jet power and accretion disk luminosity \citep{Falcke1995,Falcke1999}. Adding to the handful of spatially resolved black hole X-ray binary jets in the hard state is critical towards refining physical models for disk/jet coupling.

\subsection{The candidate black hole X-ray binary AT2019wey}

AT2019wey was first discovered as an optical transient by the Asteroid Terrestrial-impact Last Alert System (ATLAS) on December 7, 2019 \citep{Tonry2019}. A few months later on March 18, 2020, Spektrum-Roentgen-Gamma (SRG) discovered an X-ray source consistent with the position of the ATLAS detection and classified it as a hostless transient or potential supernova \citep{Mereminskiy2020}, while \citet{BLLacATel} suggested it may be a BL Lac type object. However, the discovery of hydrogen absorption lines with a redshift $z=0$ led \citet{YaoXrayAtel} to posit a Galactic accreting binary origin.

Several physical constraints on AT2019wey were derived by \citet{Yao2020MWL} through a multi-wavelength follow-up campaign. The distance to AT2019wey was constrained to be between 1--10 kpc. The lower limit of 1\,kpc was derived from the amount of observed extinction, $0.8 \lesssim E(B-V) \lesssim 1.2$ mag, calculated from Na\,I D absorption lines in the optical spectrum. The Galactic anticenter sightline of AT2019wey led to an upper distance limit of 10 kpc. A combination of extinction measurements and historical optical observations at the location of AT2019wey also constrain the mass of the companion star to be $\lesssim 0.8 \Msun$. On August 2, 2020, VLA observations of AT2019wey revealed an optically thin spectrum from 1--12\,GHz, which stood in contrast with the previous measurements of an optically thick spectrum on May 27, 2020 \citep{YaoVLAAtel,JVLAATel}. Subsequent radio spectra taken on August 14, 21, and 28 showed a return to an optically thick spectrum; however, the flux density continued to increase. The compact object was determined to be a candidate black hole by comparing the radio and optical luminosities of the system to other known binaries with comparable X-ray luminosities. The radio and optical luminosities were found to be well above that expected for a neutron star accretor across the entire estimated distance range. 

\citet{Yao2020Xray} present a detailed X-ray observational investigation of AT2019wey. X-ray observations from five telescopes -- the Neutron Star Interior Composition Explorer (NICER), Nuclear Spectroscopic Telescope ARray (NuSTAR), the Chandra X-ray Observatory, the Neil Gehrels Swift Observatory, and the Monitor of All-sky X-ray Image (MAXI) -- were used to monitor the flux and spectrum of the source. These data show that AT2019wey is in a hard state throughout its entire period of activity, although the spectrum softens between August 21, 2020 and September 28, 2020. The X-ray photon index is observed to steepen from $1.7 < \Gamma < 2.0$ to $2.0 < \Gamma < 2.3$; however, it is never seen to transition into a fully soft state. Additionally, using a model fit to the reflection spectrum observed by NICER and NuSTAR and paying close attention to the residuals around the Fe line, \citet{Yao2020Xray} infer that the inclination angle of the system must be $i\lesssim 30\degree$.

We present observations of AT2019wey with the Very Long Baseline Array (VLBA) to attempt to resolve the rapidly evolving radio source. In section~\ref{sec:obs}, we provide details on our VLBA observation as well as our data analysis procedure and in section~\ref{sec:results} we summarize the observed properties of AT2019wey. In section~\ref{sec:discussion}, we discuss parallels between AT2019wey and MAXI\,J1836$-$194 and provides an analysis of the minimum energy and power of the system.

\begin{figure*}[h]
\centering
\includegraphics[width=0.8\textwidth]{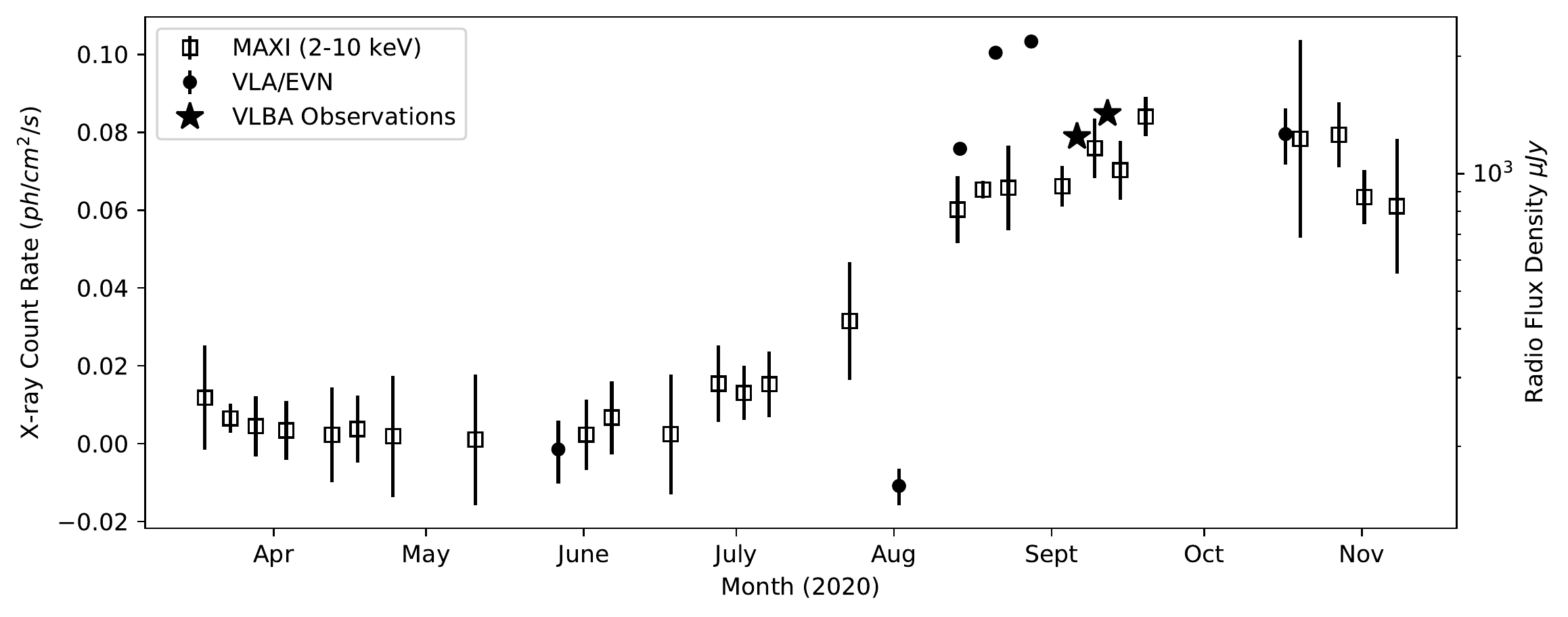}
\caption{\label{fig:lightcurve} Lightcurves showing radio observations (scaled to 4.8\,GHz) and X-ray observations from MAXI (2-10\,keV). The stars on the plot show the observations discussed in this work.}
\end{figure*}

\begin{figure*}[t]
\centering
\includegraphics[width=0.46\textwidth]{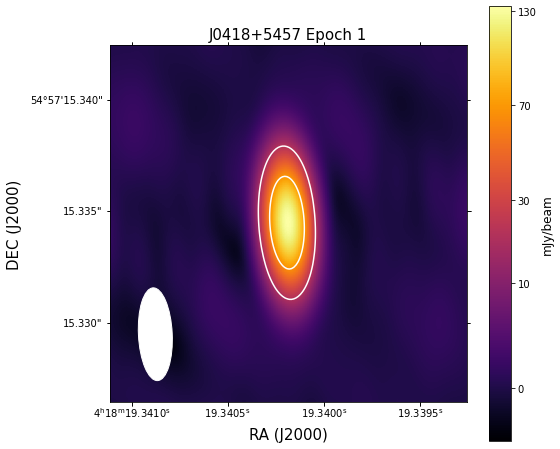}
\includegraphics[width=0.45\textwidth]{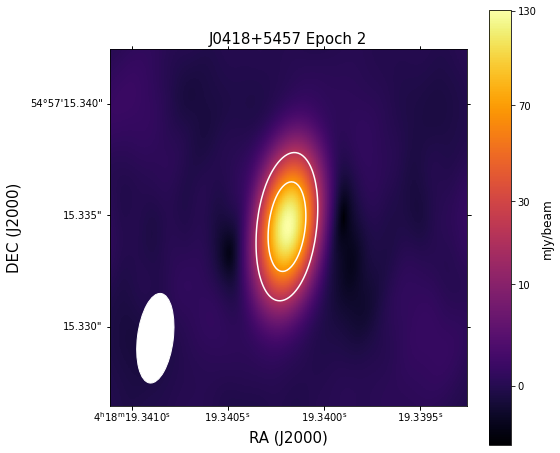}
\includegraphics[width=0.45\textwidth]{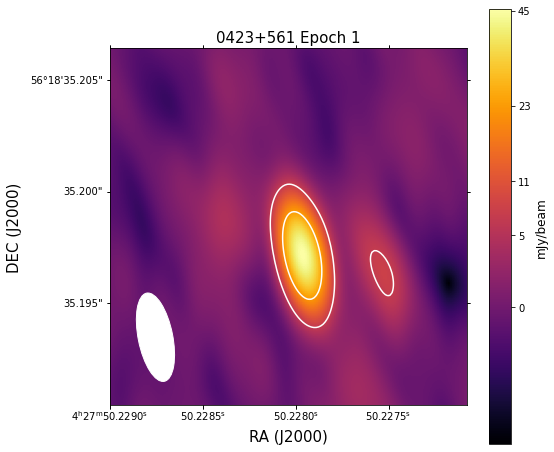}
\includegraphics[width=0.45\textwidth]{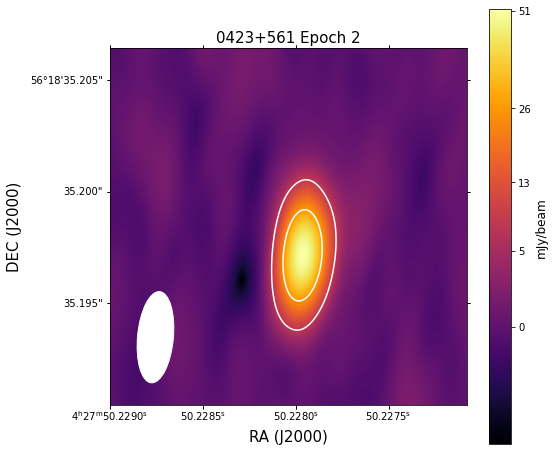}
\caption{\label{fig:phasechecks}CLEAN images of the phase reference (top row), J0418+5457, and the check source (bottom row), 0423+561, for both epochs. The images are consistent with point sources. Also depicted on the images are 15\% and 50\% flux density contours.}
\end{figure*}

\begin{figure*}[t]
\centering
\includegraphics[width=0.45\textwidth]{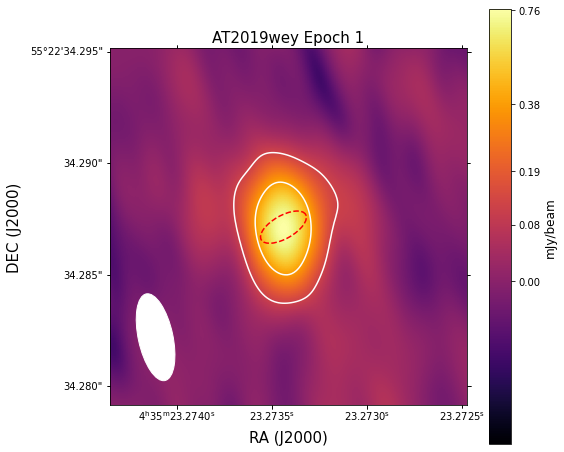}
\includegraphics[width=0.45\textwidth]{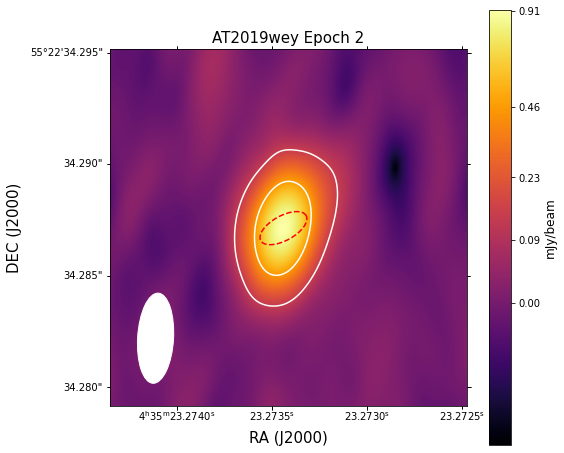}
\caption{\label{fig:2019weyimg}CLEAN images of AT2019wey for both epochs with 15\% and 50\% flux density contours. The red dotted line represents the single best fit deconvolved ellipse across both epochs for the source.}
\end{figure*}

\section{Observation and Analysis Procedures} \label{sec:obs}

Fig.~\ref{fig:lightcurve} shows X-ray and radio light curves for AT2019wey. The X-ray data are taken by the Monitor of All-sky X-ray Image (MAXI) telescope in the 2-10\,keV band \citep{MAXI}. The radio data point depicted on May 27 was taken with the VLA at 6.0\,GHz while the observations between August 2 and August 28 were taken with the VLA at 3.5\,GHz. These measurements were scaled to 4.8\,GHz using the spectral indices published by \cite{Yao2020MWL}. The radio data point shown on October 17 was taken by \cite{EVNATel} with the European VLBI Network (EVN) at 6.7\,GHz; this point was scaled to 4.8\,GHz on the light curve as well using the spectral index from \cite{Yao2020MWL} on August 28. 

The two epochs of observations of AT2019wey at 4.8\,GHz were obtained with the VLBA on September 6 and September 12 and were processed with the DiFX correlator \citep{DifX}. Both three-hour epochs were phase referenced, with alternating scans of 3.5 minutes on AT2019wey and 40 seconds on the phase reference (J0418+5457). The phase reference is at an assumed location of RA = $04^h18^m19.3401920^{s}$ and Dec = $54\degree57'15.334490"$, an angular separation of $2.47\degree$ from the target, and was chosen because of its inclusion in the third realization of the International Celestial Reference Frame (ICRF3) \citep{Charlot2020}. Each epoch also contained 4-minute observations of a check source (J0427+5618), and 6-minute observations of a bandpass calibrator (J0555+3948). 

Calibration and imaging for the observations were carried out in AIPS \citep{Greisen2003} using standard procedures. To derive phase solutions, we performed global fringe fitting on the phase calibrator followed by two rounds of phase-only self-calibration and one round of amplitude+phase self-calibration. The self-calibration was done with two-minute solution intervals and assumed a point source model with the catalog flux density for the phase calibrator. Once phase variations of less than \(\pm 5\degree\) were reached for all stations, the phase solutions were applied and no further self-calibration was performed on either the check source or AT2019wey. All images were created with natural weighting to maximize sensitivity. Images of the phase calibrator and check source are shown in Fig.~\ref{fig:phasechecks} and images of AT2019wey are shown in Fig.~\ref{fig:2019weyimg}. The full-width half maximum of the Gaussian model used to approximate the synthesized beam (shown in corners of the images) is approximately a 4\,mas $\times$ 1.5\,mas ellipse in both epochs. The estimated deconvolved component for AT2019wey for each epoch is represented by the red dotted ellipse in Fig.~\ref{fig:2019weyimg}.

\begin{figure*}[t]
\centering
\includegraphics[width=0.5\textwidth]{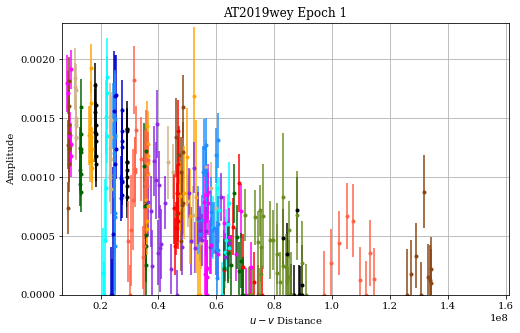}
\includegraphics[width=0.5\textwidth]{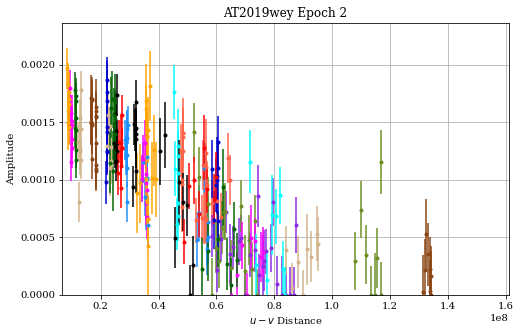}
\caption{\label{fig:2019weyuv}UV-amplitudes of AT2019wey for both epochs, with colorized points representing different baseline pairs.}
\end{figure*}

\begin{deluxetable*}{cccccc}
\tabletypesize{\footnotesize}
\tablewidth{0pt}

\tablecaption{Properties for AT2019wey, approximating the source structure as a 2D Gaussian. Though this geometry does not reflect the true source structure, more complex features can not be extracted from these observations.}
\label{tab:source_properties}
 
\tablehead{
\colhead{RA (J2000)} & \colhead{Dec (J2000)} & \colhead{Flux Density} & \colhead{Major Axis} & \colhead{Minor Axis} & \colhead{Position Angle}
}

\startdata
$04^h 35^m 23.27345^s \pm 0.00028^s$ & $55\degree 22' 34.28715" \pm 0.00017"$ & \(1.35 \pm 0.02\) mJy & \(2.13 \pm 0.10\) mas & \(0.80 \pm 0.18\) mas & \(122\degree \pm 4\degree \)
\enddata
\end{deluxetable*}

\begin{figure*}[t]
\centering
\includegraphics[width=0.45\textwidth]{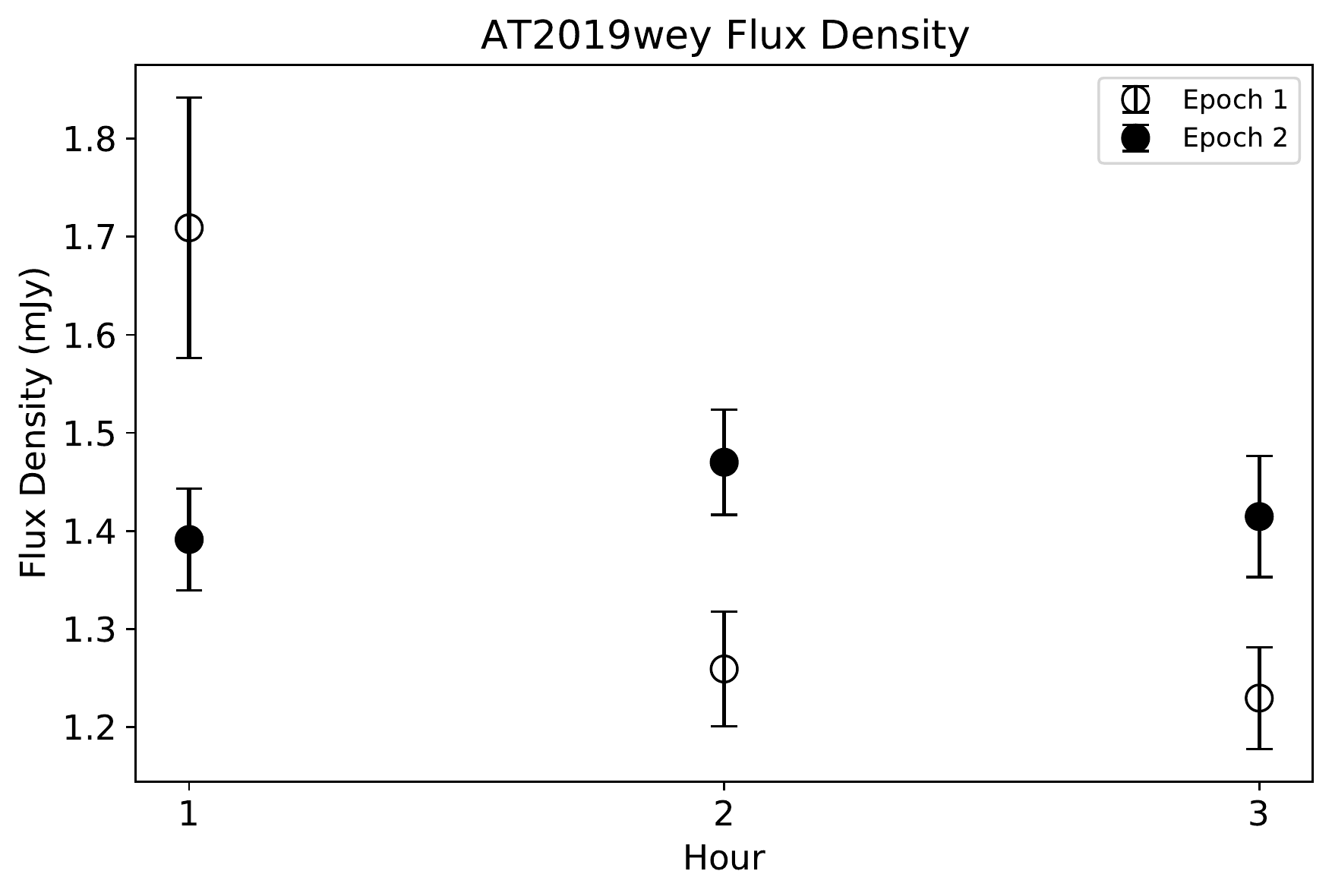}
\includegraphics[width=0.45\textwidth]{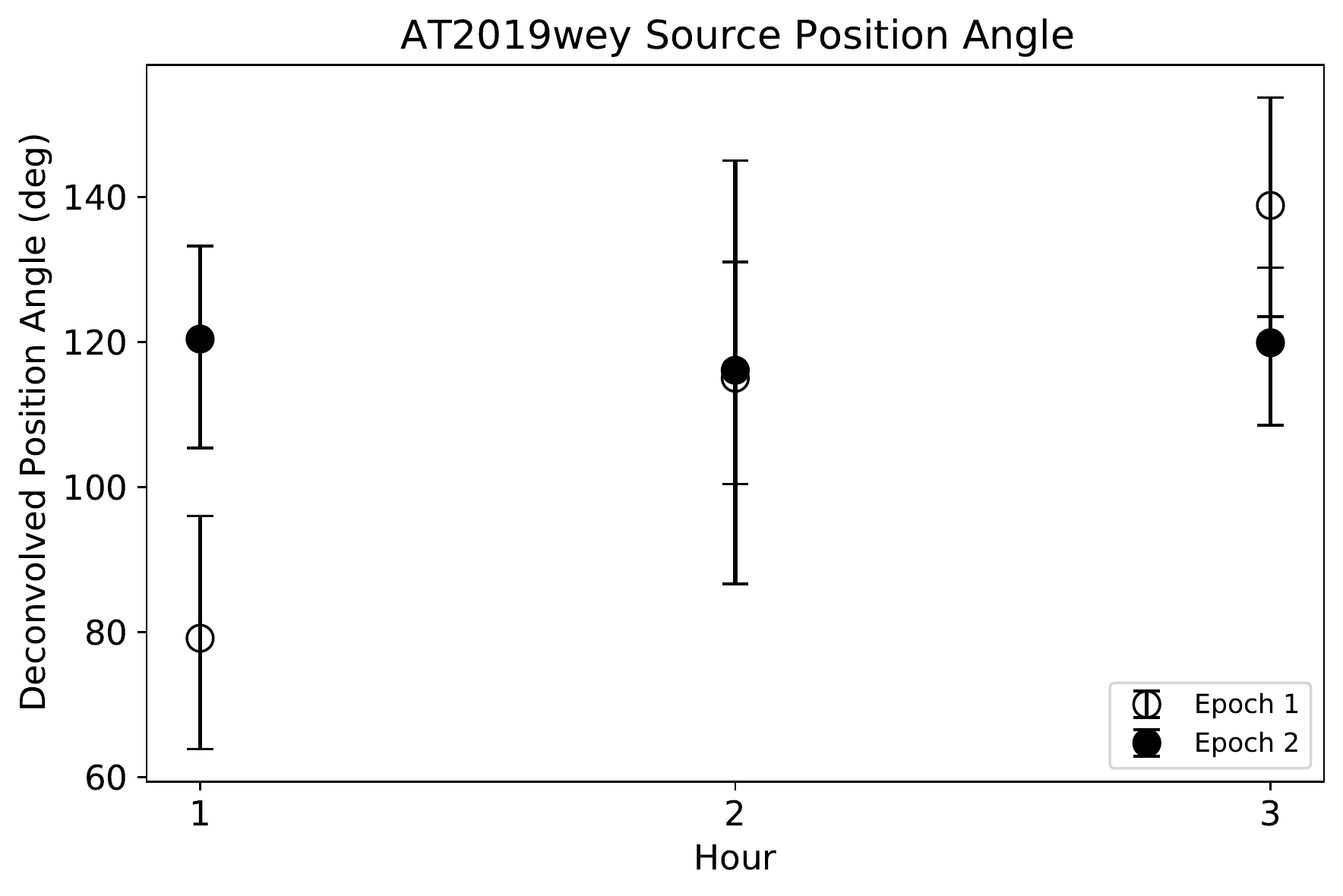}
\includegraphics[width=0.45\textwidth]{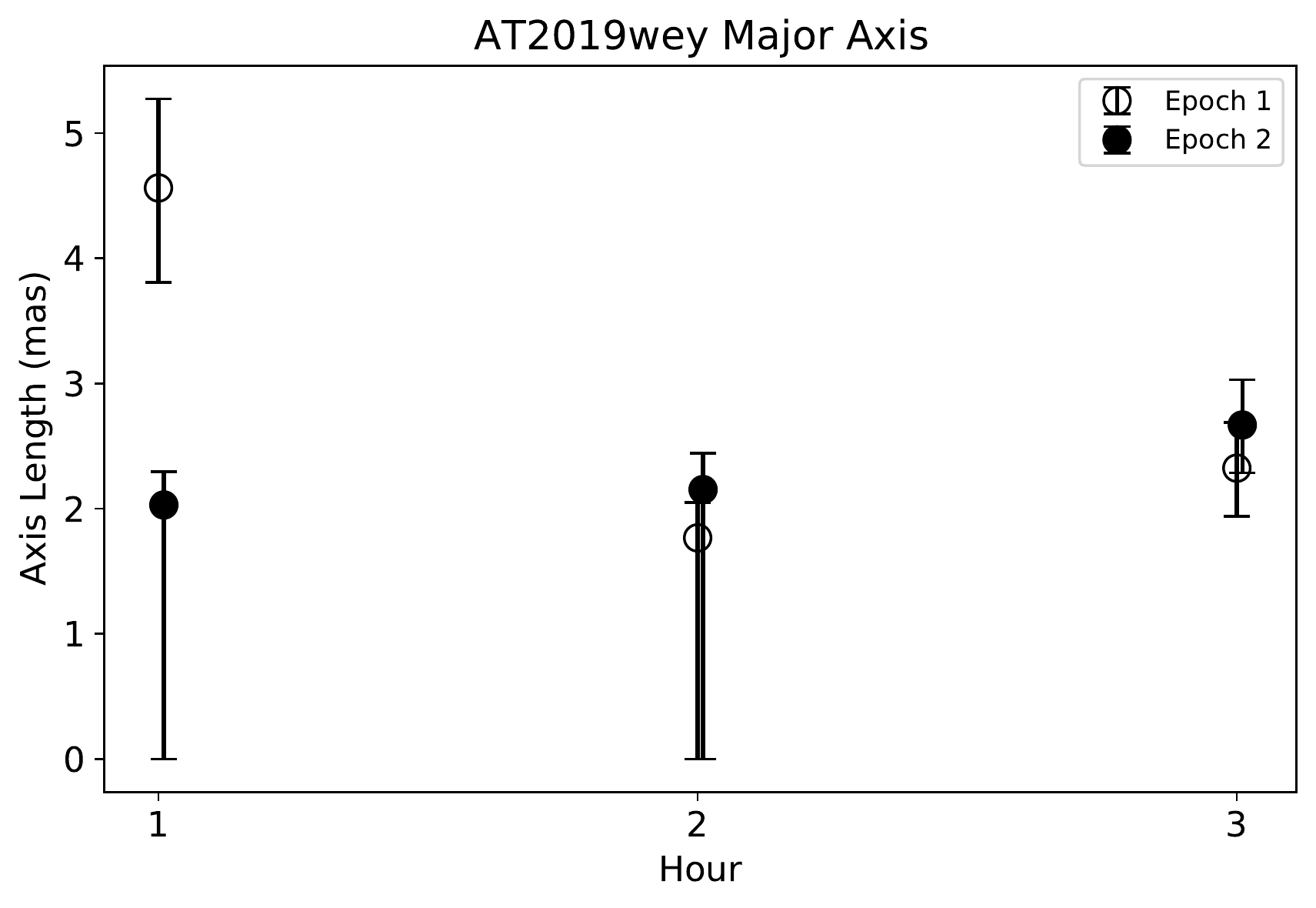}
\includegraphics[width=0.45\textwidth]{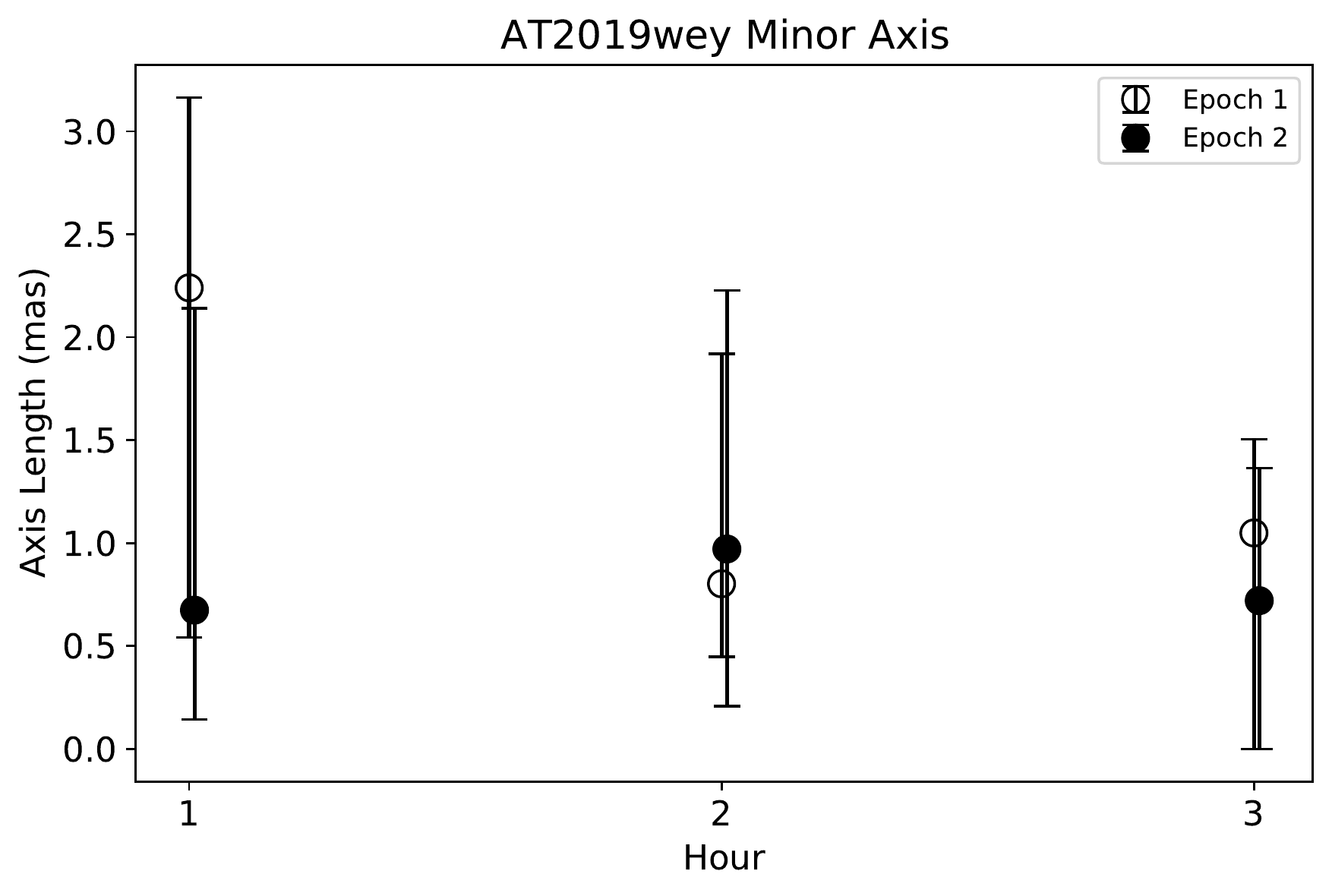}
\caption{\label{fig:2019weyjmfit}Model parameters derived for a Gaussian image plane fit to hour long observation blocks of AT2019wey. All of the geometric parameters are estimations of a deconvolved component. Top left: flux density, top right: position angle measured east of north, bottom left: major axis, bottom right: minor axis}
\end{figure*}
\section{Results} \label{sec:results}
The images and uv-amplitudes of AT2019wey both show that the source is resolved. The images (Fig.~\ref{fig:2019weyimg}) show a source that is clearly wider and oriented at a different position angle than the elliptical model of the synthesized beam. Plots of the uv-amplitudes, coherently averaged for 20 minute intervals, are shown in Fig.~\ref{fig:2019weyuv} along with uncertainties. The amplitudes are clearly not constant as a function of uv-distance, as would be expected for a point source. 

We attempted to fit a 2-D Gaussian component to the visibilities using the \texttt{uvfit} task in AIPS; however, low signal to noise in individual visibility measurements yielded unreliable results for a 2-D Gaussian model fit to visibilities. 
Instead, we performed an image plane fit using the AIPS task \texttt{jmfit}, which uses a least squares approach to fit a 2-D Gaussian component to the image and then deconvolves the CLEAN beam from the fitted component to estimate a 2-D Gaussian model for the true source geometry.  The \texttt{jmfit} task also provides uncertainties for all of these parameters; however, the task will report the lower limits of the deconvolved major and minor axis as 0\,mas if a reliable uncertainty cannot be derived. As we are confident that the source is resolved, in these cases we assume symmetric uncertainties based on the estimate of the upper limit for use in any calculations. To achieve convergence with the \texttt{jmfit} task, we fix the position of the 2-D Gaussian to the position of the phase center of the image. The errors on the RA and Dec of the source position are then dominated by the cataloged position errors of the phase reference.

To investigate time variability in the phase calibrator and AT2019wey, we image and perform model fitting with \texttt{jmfit} for three hour-long blocks in each epoch. The fitted flux density, position angle reported east of north, the major axis, and the minor axis for each hour of both epoch are shown along with uncertainties in Fig.~\ref{fig:2019weyjmfit}. The mean best fit values and uncertainties are summarized in Table~\ref{tab:source_properties}. We see that AT2019wey has a flux density of $1.35 \pm 0.02$\,mJy and an approximated 2-D Gaussian geometry with a major and minor axis of $2.13 \pm 0.10$\,mas and $0.80 \pm 0.18$\,mas respectively at a position angle of $122\degree \pm 4\degree$. These values closely agree with those derived from performing a model fit on data combining the two epochs. Though the formal uncertainties on these parameters are small, this does not imply the true source geometry is well modeled by a Gaussian; however, subtracting this model from the visibilities and imaging the residuals reveals only noise, an indication that it would be difficult to extract more complex structure from this data. The model does illustrate relatively stable source properties and facilitates a rough estimate of source power.

Although the source geometries of AT2019wey remain relatively consistent within the uncertainties across both epochs, there are some notable deviations. The best fit parameters are anomalous for the first hour of the first epoch. Inspection of the phase calibrator showed no significant phase or gain fluctuations during this time, indicating that the anomalous measurement is likely not caused by calibration error. As a light-crossing time of one hour is equivalent to a distance of several AU, a scale similar to the synthesized beam-width for low distance estimates, it cannot be ruled out that this measurement is due to true variability in the X-ray binary. The fitted flux density is also around 15\% higher in the second epoch than the first. As the source is seen to be fading in the radio lightcurve presented in \cite{Yao2020MWL}, we do not expect the second epoch to show significant increase in brightness. A similar level of variation is seen in the flux density of the check source, which was fitted to 45.6\,mJy in epoch 1 compared to 54.9\,mJy in epoch 2. This may indicate residual phase error in the first epoch leading to a lower flux density measurement, but intrinsic source variability on these small angular scales cannot be ruled out. The use of dynamical imaging and attempts to super-resolve the source, left to a later work, would provide a way to verify potential short timescale evolution.

\section{Discussion} \label{sec:discussion}


A comparison of these VLBA observations of AT2019wey and observations taken of similar systems, specifically GRS\,1915+105 and MAXI\,J1836$-$194, provides compelling evidence that the radio source we observe is likely a steady compact jet. Compact radio sources were observed while both of these systems were in hard X-ray spectral states. \citet{Dhawan2000} report a well-resolved, elongated radio source in GRS\,1915+105 that exhibits a shallow radio spectral index, $\alpha \lesssim 0.5$, and a steady position angle over two years of observations. Although \citet{Russell2015} report only a marginally-resolved radio source in MAXI\,J1836$-$194 with a steeper spectral index of $\alpha \lesssim 0.8$, the source also shows a stable position angle over two months of observations. The fact that AT2019wey was observed in the hard X-ray state with a nearly flat radio spectral index of $\alpha \sim 0.2$ and the position angle of VLBA source remains stable between the two epochs of observation indicate the presence of compact steady jet.

The measurement of an angular size enables us to estimate the total energy in the source following standard synchrotron theory  \citep[following][]{Pacholczyk1970}. We assume an ellipsoidal structure for the source, with a projected shape corresponding to the 2-D Gaussian described above. At a fiducial distance of $D=3$\,kpc, we adopt a source volume of $V=10^{42}$\,cm$^{-3}$. We assume a flat radio spectrum between 1--12\,GHz only (spanning the VLA observations of AT2019wey), with a flux density of 1.35\,mJy. The minimum energy required to power the synchrotron source (relativistic particles and magnetic fields) is then 
\begin{equation}
    E_{\rm min} \approx 5\times10^{38}\left(\frac{D}{\rm 3\,kpc}\right)^{\frac{17}{7}}\left(\frac{V}{\rm 10^{42}\,cm^{3}}\right)^{\frac{3}{7}}\,{\rm erg}. 
\end{equation}
The corresponding mean magnetic field strength is $\sim0.07$\,G, implying a relativistic-lepton Lorentz factor of $\gamma\approx250$ for 12\,GHz emission. Assuming a characteristic particle acceleration timescale corresponding to the light crossing time of the source of $\sim3\times10^{3}$\,s at 3\,kpc, the power dissipation in the source is approximately 
\begin{equation}
    P \gtrsim 2\times10^{35}\left(\frac{D}{\rm 3\,kpc}\right)^{\frac{10}{7}}\left(\frac{V}{\rm 10^{42}\,cm^{3}}\right)^{\frac{3}{7}}\,{\rm erg\,s^{-1}}. 
\end{equation}
We emphasize that this is a lower limit given the limited band used to calculate the total radio luminosity, and the minimum-energy assumption. 

The inferred power is remarkably close to the $\sim10^{36}$\,erg\,s$^{-1}$ X-ray luminosity of AT2019wey found by \citet{Yao2020Xray} for a 3\,kpc distance. The luminosity of the thermal emission from the disk is likely a few tens of percent of this total. This correspondance has been observed previously in the hard and plateau states of the GRS\,1915+105 \citep[e.g.,][]{Dhawan2000}, and is a critical assumption of models for symbiotic disk-jet systems \citep{Falcke1999}. Although we resolve the radio source in AT2019wey, we have no compelling morphological evidence for a jet. Nonetheless, the panchromatic properties of AT2019wey are closely similar to low-mass black hole X-ray binaries in which relativistic jets have been observed \citep{Yao2020MWL}. We therefore interpret the resolved source as a compact steady jet, with a power that is comparable to the accretion-disk X-ray luminosity. 

%
%
\section{Conclusion} \label{sec:conclude}
We present here two epochs of 4.8\,GHz VLBA observations of candidate black hole low-mass X-ray binary system AT2019wey following a period of X-ray and radio brightening. The observations revealed a resolved source with deconvolved source geometries that are relatively constant across both epochs. Together with the observed X-ray spectrum, we interpret these results to indicate the presence of a compact, steady jet. Using the angular scale derived from image plane fits of a 2-D Gaussian component to the source, we show that the power dissipation from the jet is comparable to the X-ray luminosity, consistent with a standard assumption of models for disk/jet coupling. 

Thus far, spatially resolved compact jets in X-ray binaries in the hard spectral state have only been observed for five systems, including AT2019wey. The next-generation Very Large Array (ngVLA) and the Square Kilometre Array (SKA) are ideal instruments to expand on this limited sample \citep{fender15,MaccaronengVLA}. With their combination of long baselines and extreme sensitivity, the ngVLA and SKA will enable high-cadence monitoring of the flux densities and multi-scale morphologies of an extended sample of X-ray binaries. The ngVLA may in fact prove to be a discovery engine for accretion Galactic black holes through astrometric surveys \citep{Maccarone2}. Observations of these systems during a variety of states and state transitions are required to broaden our understanding of the connections between accretion states and jets. 


\section{Acknowledgements} \label{sec:acknowledge}
We thank Tim Pearson, Gregg Hallinan, and Katie Bouman for useful discussions on interpreting these results. These observations were conducted with the Very Long Baseline Array. The National Radio Astronomy Observatory is a facility of the National Science Foundation operated under cooperative agreement by Associated Universities, Inc. This research has made use of NASA's Astrophysics Data System.

\bibliography{2019weyVLBA}{}
\bibliographystyle{aasjournal}

\end{document}